\documentclass[]{iopart}
\usepackage{amssymb}
\usepackage{epsf}
\usepackage{harvard}
\begin{document}
\jl{1}  

\newcommand{\fig}[2]{\epsfxsize=#1\epsfbox{#2}}
\title[Renormalization of two dimensionnal modular invariant Coulomb gas]
{Renormalization of 
modular invariant Coulomb gas and Sine-Gordon theories, and quantum Hall flow diagram
}
\author{David Carpentier}
\address{ 
Laboratoire de Physique Th{\'e}orique de l'Ecole Normale Sup{\'e}rieure\\
24, Rue Lhomond 75005 Paris}

%
%
%

%
\begin{abstract}
Using the renormalisation group (RG) we study two dimensional 
electromagnetic coulomb gas and extended Sine-Gordon theories 
invariant under the modular group 
$SL(2,\mathbb{Z})$. The flow diagram is established from the scaling
equations, and we
derive the critical behaviour at the various transition points of the
diagram. Following proposal for a $SL(2,\mathbb{Z})$ duality between different
quantum Hall fluids, we discuss 
the analogy between this flow and the global quantum Hall phase diagram.  
\end{abstract}

\section{Introduction}

 In statistical physics, self-duality in the sense of Kramers and Wannier maps 
the high temperature regime of a given model with its low temperature regime. 
 In two dimensions, examples of self-dual theories are provided by the  
2 (Ising),3 and 4 states Potts, the Ashkin Teller, and the clock models. All of
them can be represented in terms of an electromagnetic  Coulomb
gas \cite{kadanoff78,kadanoff79}, allowing to understand this self-duality 
as an exchange of the electric and magnetic components of the
charges of the equivalent Coulomb gas \cite{kadanoff78}.   
 This self-duality  allows to locate the transition point, whose 
study requires however 
a full renormalization group study, either directly on the Coulomb gas
formulation \cite{nienhuis87} or on the associated Sine-Gordon theory
\cite{boyanovsky89}.

 More than fifteen years ago, Cardy uncovered a generalization of this simple
duality by adding a topological coupling between magnetic and electric charges
of a two flavors Coulomb gas \cite{cardy82b}.  Motivated by the study of
oblique confinement in four dimensional $\mathbb{Z}_{p}$ lattice gauge theory with a
topological $\theta$ term, \citeasnoun{cardy82a} formulated a Coulomb gas
where the presence of the $\theta$ term considerably enlarge the usual Kramers
Wannier duality $g\to 1/g$
to the full modular group 
$SL (2,\mathbb{Z})$ \cite{cardy82b}. This $\theta$ coupling was later extended to
higher dimensions in \cite{shapere89}. 
 By analogy with the work of Kramers and Wannier, 
\citeasnoun{cardy82b} derived the location of the numerous transition points
in the presence of this topological coupling, as the invariant points of the
modular group.  

 The purpose of this article is to explicitly study using the renormalization
group the scaling behaviour of an extension of the Cardy-Rabinovici
model. Besides the critical behaviour associated with the transition points
identified by Cardy, this renormalization study allows to find the scaling
flow of the model : the fixed points defining different phases are defined and
related to the transitions points. We thus find a condition that two different
phases must fullfill to be related by a transition. This is, to our
knowledge, the first explicit renormalization of a two dimensional modular
invariant model.     


 Besides the applications of this extended Coulomb gas to the previously cited
statistical models in two dimensions and to fermion models in 1+1 dimension
via its Sine-Gordon representation, 
such two dimensional modular invariant
theories are of interest for the study of the global phase diagram of the
quantum Hall effect. 
 This effect correspond to the quantization of the transverse conductivity
$\sigma_{xy}$ of a two dimensional gas of electrons in a strong transverse
magnetic field, and it is now well understood in terms of the microscopic
Laughlin waves functions \cite{laughlin83}. The problem of the nature of the
transitions between 
the different quantum Hall fluids is of particular interest and remain
unsolved.  
 In particular the notion of superuniversality was proposed to explain the
similar behaviour of all the transitions in the global phase
diagram of the quantum Hall effect \cite{kivelson92}. Such a superuniversality
can be deduced 
from a duality of the underlying model, relating all the transition points with
each other. Indeed  
the similarity between the phase diagram of
\cite{cardy82b} and the expected renormalization flow diagram in the two
parameters scaling model of the Quantum Hall effect was noticed
\cite{shapere89} before this proposal. Later on \citeasnoun{lutken92} related
more precisely the properties of this phase diagram with the presence of a $SL
(2,\mathbb{Z})$ symmetry, in the framework of a two parameters
$\sigma_{xx},\sigma_{xy}$ scaling theory. 

 On the other hand, the experiments of \cite{shahar96} on the transition
between the $\nu=\frac{1}{3}$ and the Hall insulator have shown a reflection 
relation between the nonlinear current density and electric field on both side
of the transition. Within
the effective description of the quantum Hall effect
, this reflection was interpreted as a duality which
exchanges the electric and magnetic component of the low lying
excitations \cite{shahar96}. Motivated by these experimental results, several
authors focused on modular invariant models \cite{fradkin96,pryadko96}.  
 Even though no derivation of a microscopic modular invariant model exists,
these studies focused on the general constraint from the modular symmetry 
on the phase diagram and expressions for the conductivity at the
transitions \cite{lutken92,dolan98}.  
 The phase diagram of the modular invariant Coulomb gas we study in this paper
is expected to mimic the one of the quantum Hall effect. As 
an explicit renormalization of a modular invariant model in
two dimension, we expect this study to be helpfull for the description of the
critical behaviour of the quantum Hall transitions.     

 In another context, it is interesting to notice that  
modular invariant Coulomb gas also appeared in the work of 
\citeasnoun{callan91} on the dissipative motion of a
charged particle in two 
dimension in a transversed magnetic field   and a periodic electric
 potential. Using a mapping to a one dimensional version of the model of
\cite{cardy82a}, these authors found a 
phase diagram as a function of the dissipation strength and the magnetic field which
resembles the diagram obtained from our renormalization study.  

 This article is organized as follows : 
in the section \ref{part:SG} we define the model both in its Sine-Gordon
version and as a lattice Villain model with topological $\theta$ term. The
mapping to an electromagnetic Coulomb gas is then obtained (\ref{part:CG})
together with the symmetry $SL(2,\mathbb{Z})$ (\ref{part:symmetrie}). In the
next section (\ref{part:RGeqs}) we derive the general renormalization group
equations for the model using a Kosterlitz scheme, and the phase diagram is
established in part \ref{part:analyse} together with the critical behaviour
at the critical points of this diagram. 

\section{Electromagnetic Coulomb gas}\label{part:model}

\subsection{Extended Sine-Gordon model and theta terms}\label{part:SG}
 In this article we will consider the two dimensional 
extended Sine-Gordon model defined in
its more general form by the partition function 
\[
Z=\prod_{a}\int d[\phi^{a}] e^{{\cal A}[\phi^{a},\tilde{\phi}^{a}]}
\]
 with the action 
\begin{eqnarray}\nonumber 
\fl 
{\cal A}[\phi^{a},\tilde{\phi}^{a}]=\int \frac{d^{2}{\bf r}}{a^{2}}
\left(-\frac{1}{4 \pi}\sum_{a,b=1}^{2k} 
[g^{-1}]^{ab}\partial_{i} \phi^{a}\partial_{i} \phi^{b}
\right. \\ \left. 
\label{H-general}
+2y\sum_{a}  \cos (\phi^{a}) 
+ 2\tilde{y} \sum_{a} \cos p
\left([g^{-1}]^{ab}\tilde{\phi}^{b}-M^{ab}\phi^{b}\right)  \right)
\end{eqnarray}
 In this definition we consider $2k$ integer valued fields $\phi^{a}$ and
their dual fields
 $\tilde{\phi}^{a}$ defined by
$i \partial_{i}\tilde{\phi}=\epsilon_{ij}\partial_{j}\phi$, where $i=1,2$
labels the two directions of the plane, and $\epsilon_{ij}$ is the
antisymmetric tensor.  
The coupling matrix $g^{ab}$
and $M^{ab}$ are $2k\times 2k$ matrices;  $M^{ab}$ has to be antisymmetric 
for renormalisability of the model (see the following) : $M^{ab}=-M^{ba}$.  
 In this context the operators 
\begin{equation}\label{parafermions}
{\cal O}_{{\bf n,m}} ({\bf r})=\exp ~
i\left(\sum_{a}n_{a}\phi^{a}({\bf r})
+\sum_{a}m_{a}\tilde{\phi}^{a}({\bf r})\right) 
\end{equation}
correspond to the parafermions operators of \citeasnoun{fradkin80} which
create a vector electric charge ${\bf n}$ and magnetic charge ${\bf m}$ in
site ${\bf r}$.  

 Alternatively we can consider a $\mathbb{Z}_{p}$ gauge theory defined on the
square lattice by $2k$ Villain models perturbed by symmetry breaking fields of
strength $y$ and coupled by a (topological) term  :   
\begin{eqnarray}\nonumber \fl
{\cal A}_{villain}=-\frac{1}{4\pi}\sum_{\alpha}\sum_{i=1,2} 
[g^{-1}]^{ab}
\left(\partial_{i}\phi_{\alpha}^{a}-2\pi p A_{\alpha,i}^{a} \right)
\left(\partial_{i}\phi_{\alpha}^{b}-2\pi p A_{\alpha,i}^{b} \right)
+\ln (y)\sum _{\alpha}n_{\alpha}^{a}n_{\alpha}^{a}\\\label{villain} 
+i \sum_{\alpha}n^{a}_{\alpha}\phi^{a}_{\alpha}
-i \frac{1}{4 \pi}\sum_{\alpha}M^{ab}\epsilon_{ij}
\left(\partial_{i}\phi_{\alpha}^{a}-2\pi  p A_{\alpha,i}^{a} \right)
\left(\partial_{j}\phi_{\alpha}^{b}-2\pi  p A_{\alpha,j}^{b} \right)
\end{eqnarray}
 where the $A_{\alpha,i}^{a}$ correspond to the integer valued gauge field
defined on the bonds of the square lattice, and $\partial_{x}\phi^{a}_{\alpha}$ 
 is the discrete derivative
$\phi^{a}_{\alpha+x}-\phi_{\alpha}^{a}$. 
 Here and in the following summation over repeated indices is assumed. 
 
The meaning of the coupling between the different fields $\phi^{a}$ becomes
clearer upon restriction to a two components model ($k=1$) with the coupling
constants 
\begin{equation}\label{model-p2}
g^{ac}=g\delta^{ac} \quad ;\quad  M^{ac}=\frac{\theta}{2\pi}\epsilon^{ac}
\quad \quad \quad \quad  (k=1)
\end{equation}
 By considering the 4 component fields $\phi_{\mu}=(0,0,\phi^{1},\phi^{2})$
which depends only on $x_{1},x_{2}$  we can write the quadratic term in
(\ref{villain}) as $F_{\mu\nu}F^{\mu\nu}$ where $\mu,\nu=1,\dots 4$ and
$F_{\mu\nu}$ is the usual electromagnetic tensor associated with the field
$\phi^{\mu}$ :
$F_{\mu\nu}=\partial_{\mu}\phi_{\nu}-\partial_{\nu}\phi_{\mu}- 2 \pi p A_{\mu\nu}$. the
second term can be interpreted as a coupling between electric particle and the
field $\phi_{\mu}$ while the last term reads
$(\theta/2\pi)\epsilon_{\mu\nu\rho\sigma}F^{\mu\nu}F^{\rho\sigma}=
(\theta/2\pi)F^{\mu\nu}\tilde{F}_{\mu\nu}$,
which is known as a topological $\theta$ term. This model and its Coulomb gas
formulation in two dimensions was first studied by
\citeasnoun{cardy82a}. 
The general $M$ coupling in (\ref{H-general}) is thus 
the analog of this topological coupling for the $2k$ real components field in two
dimensions. Other extensions to higher dimensions of space were developped 
in \cite{shapere89}.  

 Although in the following, 
we will derive renormalization group equation for the general model
(\ref{H-general}),  
we will only analyze in details the scaling behaviour of the simpler
model (\ref{model-p2}) 
with only electric charges in the first component ($a=1$) and magnetic in the
second (see below). This model is defined by the following restriction
of (\ref{H-general}) :  
\begin{equation} \label{model-specific}
\fl
H=\frac{1}{4\pi g}\int \frac{d^{2}{\bf r}}{a^{2}}
\left( (\partial \phi^{1})^{2}+(\partial \phi^{2})^{2}\right)
+y \cos (\phi^{1}) 
+ \tilde{y} \cos p
\left(g^{-1} \tilde{\phi}^{2}-\frac{\theta}{2\pi}\phi^{1}\right) 
\end{equation}
 We can now express the partition function of the above models in terms of
electromagnetic Coulomb gas, extending the usual case of
\cite{nienhuis87}.

\subsection{$SL(2,\mathbb{Z})$ invariant electromagnetic Coulomb gas}\label{part:CG}

 To obtain the Coulomb gas formulation of the above model, we first use the
Villain approximation \cite{villain75} 
of the cosine coupling in (\ref{H-general}) : 
\[
e^{2y\cos \phi}\sim \sum_{n=0,\pm 1} e^{i n.\phi+n^{2} \ln (y)}
\]
 This approximation is valid for small coupling strength $y$. 
The models with both forms of the interaction are known to be in the
same universality class without the topological term \cite{boyanovsky89}, 
which ensures the same in 
our case. Within this
approximation, the partition sum of (\ref{H-general}) 
consist of a trace over the fields $\phi^{a}({\bf r})$ and the electric and 
magnetic  
charge density $n^{a} ({\bf r})$ and ${\bf m}^{a} ({\bf r})$ of the exponential
 of the action      
\begin{eqnarray*}
\fl 
\tilde{{\cal A}}= 
\int \frac{d^{2}{\bf r}}{a^{2}}\left(-\frac{1}{4\pi}\sum_{a,b} 
[g^{-1}]^{ab}\partial_{i} \phi^{a}\partial_{i} \phi^{b}
+i \sum_{a} \phi^{a} (n^{a}+p M^{ab}m^{b})+ 
i p \sum _{a} \tilde{\phi}_{a} [g^{-1}]^{ab}m_{b}\right. \\  \left. 
+\ln (y)\sum_{a} (n^{a})^{2}+\ln \tilde{y}\sum_{a} (m^{a})^{2}
\right)
\end{eqnarray*}
 In this bare model the vector charges ${\bf n} ({\bf r})$ have component
$0,\pm 1$, however upon coarse graining charges with higher components will be
generated. The lattice Villain model with a topological coupling 
(\ref{villain})  leads exactly to the same sum, where the magnetic charges 
$m^{ a} ({\bf r})$ are located on the sites of the dual lattice. 
They are defined by
the oriented  sum of the potential ${\bf  A}^{a}$ over the plaquette surrounding
the dual site : $m^{a} ({\bf r})=-\epsilon_{ij}\partial_{i}A_{j}^{a}$. 

 After integration over the field $\phi^{a} ({\bf r})$ , and using the
neutrality of the charges $\int_{{\bf r}} n^{a} ({\bf r})=\int_{{\bf r}} m^{a}
({\bf r)=0}$ (imposed by the infrared regularisation), we obtain 
an electromagnetic two dimensional coulomb gas of electric $n^{a}$
and magnetic $m^{a}$ charges, defined either on a lattice (and its dual for the
$m$ charge) or in the continuum,  
 which take value in $\mathbb{Z}^{2k}$. It is defined by the grand canonical 
partition function 
\begin{equation}\label{partition}
Z = 
\sum_{[{\bf n}_{i},{\bf m}_{i}]}' 
\left(
 \prod_{i=1}^{N}\int\frac{d^{2}{\bf r}_{i}}{a^{2}}y_{{\bf n}_{i},{\bf m}_{i}} 
\right)
e^{{\cal A}_{cg}[{\bf n,m}]}
\end{equation}
 where the $y_{{\bf n},{\bf m}}$ are the charge fugacities, and the primed sum
counts each 
distinct configuration only once. 
The corresponding action can be written
in its most general form as 
\begin{eqnarray}\nonumber
\fl {\cal A}_{cg}[{\bf n,m}]= \frac{1}{2} \left[ 
g^{ac} (n^{a}+p M^{ab} m^{b})*G*(n^{c}+p M^{cd}m^{d})
+ p^{2} (g^{-1})^{ac}m^{a}*G*m^{c}  \right]\\\label{action} 
-i~ p n^{a}*\Phi*m^{b}
\end{eqnarray}
where we assumed summation over repeated indices and used the contraction
notation : $n*G*n=\sum_{i\neq j} n_{i}G ({\bf r}_{i}-{\bf r}_{j})n_{j}$. 
In this action, the potentials $G$ and $\Phi$ are defined respectively by the
propagators 
\begin{eqnarray*}
g^{ab}G (r)=\langle \phi^{a} ({\bf 0})\phi^{b}({\bf 0})\rangle-
\langle \phi^{a} ({\bf 0})\phi^{b}({\bf r})\rangle\\
-i\Phi (r)=\langle \phi^{a} ({\bf 0})\tilde{\phi}^{b}({\bf 0})\rangle-
\langle \phi^{a} ({\bf 0})\tilde{\phi}^{b}({\bf r})\rangle
\end{eqnarray*}

 In a neutral Coulomb gas, these propagators need only to be regularised at short
distances : in the following we will use a real space hard cut-off $a$, corresponding to
hard core charges. The asymptotics of these interactions is given by 
 $G ({\bf r}) +i \Phi({\bf r}) \sim_{r\gg a}\ln (x+iy)/a$. 
 For a definition of these potentials on the lattice, see 
\cite[appendix A]{kadanoff79}.  

\subsection{Symmetries for two components charges
($k=1$)}\label{part:symmetrie} 
 We now considere more precisely the model (\ref{model-specific}). 
With the presence of the $\theta$ (or $M$) coupling, the usual Kramers-Wannier
duality $g\leftrightarrow p^{2} g^{-1}$ of the electromagnetic 
coulomb gas \cite{kadanoff78} is considerably enlarged \cite{cardy82b}. 
Besides the time reversal symmetry 
$T~ S[\{n^{a}\},\{m^{a}\},g,\theta]=S[-\{n^{a}\},\{m^{a}\},g,-\theta]$,  the $2\pi$
periodicity of the 
$\theta$ coupling translates into 
$P~S[\{n^{a}\},\{m^{a}\},g,\theta]=S[\{n^{a}-
\epsilon^{ab}. m^{b}\},\{m^{b}\},g,\theta+2\pi]$. Finally 
the action (\ref{action}) 
is also invariant under the self-dual transformation 
$D ~S[\{n^{a}\},\{m^{a}\},g,\theta]=
S[\{\epsilon^{ab}. m^{b}\},\{\epsilon^{ab}.n^{b}\},g',\theta']$ 
with   
\begin{eqnarray}\label{dual}
 \frac{g}{p^{2}} =\frac{1}{g'}+g' (\theta')^{2}~~ ;~~ 
 g\theta = -g' \theta'
\end{eqnarray}
 These transformations are better parametrised using the complex coupling
constant 
$z=p\frac{\theta}{2\pi}+ipg^{-1} $ :  the above transformation simply read 
$P (z)=z+1$ and $D
(z)=-z^{-1}$. As obvious from this writing, 
$D$ and $P$ do not commute, and they generate the whole infinite
discrete group $SL (2, {Z})$. 
 The purpose of this letter is thus to study explicitly the behaviour of the
modular invariant Coulomb gas (\ref{action}) under the renormalization group. 

\section{Renormalization {\`a} la Kosterlitz}\label{part:renormalisation}

\subsection{Renormalization group equations}\label{part:RGeqs}

 Without a $\theta$ (M) 
term, the electromagnetic Coulomb gas (\ref{action}) can be renormalized
either by following the (Anderson-Yuval) Kosterlitz scheme \cite{nienhuis87} 
or directly by an
operator product expansion in the Sine-Gordon
formulation \cite{boyanovsky89}. Both methods amounts to study the product of
the parafermion operators (\ref{parafermions}) and give the same
results. Hence for seek of simplicity, we will follow the Kosterlitz approach,
extending the classical method to the presence of the
$\theta$ term (see \cite{nienhuis87}
for a detailed review of this method).  

Upon coarse graining the model by increasing the
hard-core 
cut-off $a$, we leave the partition function (\ref{partition}) invariant by
defining scale dependent coupling constants and fugacities. Three different
contributions to these variables have to be considered  : naive rescaling,
fusion and annihilation of 
electromagnetic charges.  The naive rescaling comes simply from the change of
cut-off in the integration measure and the interaction $G (r)$ : its gives the
eigenvalue of the fugacities $y_{{\bf n},{\bf m}}$. Upon infinitesimal increase of the
cut-off $a\to \tilde{a}=ae^{dl}$, the 
distance between two neighbors charges can become less than the new
cut-off $\tilde{a}$. When these two charges form a dipole, we integrate them out
(annihilation of charges), while we simply glue them into a single charge
otherwise (fusion of charges). In this last case we get a correction to
the fugacity of the new charge, which together with the naive rescaling, reads    
 the following scaling equation for the fugacities :
\numparts
\label{RG-brut}
\begin{eqnarray}\label{rg-Y}\fl
\partial_{l} y_{{\bf n},{\bf m}} = 
\left(2-\frac{g}{2} \sum _{a}
(n^{a}+pM^{ab}.m^{b})^{2}-\frac{p^{2}}{2g}\sum_{a} (m^{a})^{2} \right)  
y_{{\bf n},{\bf m}}\\
\nonumber 
+ \pi \sum_{({\bf n}',{\bf m}')+ ({\bf n}'',{\bf m}'')= ({\bf n},{\bf m})}
\delta_{{\bf n}'.{\bf m}''+{\bf n}''.{\bf m}'}y_{{\bf n}',{\bf
m}'}y_{{\bf n}'',{\bf m}''} 
\end{eqnarray}
 with the notation ${\bf n.m}=\sum_{a}n^{a}m^{a}$.

 In the other case, the small dipoles we annihilate upon coarse graining 
screen the interaction between distant charges. In the partition
function (\ref{partition}),  the term involving two (non zero) charges 
$({\bf n}_{s},{\bf m}_{t})$
and $({\bf n}_{s},{\bf m}_{t})$ distant from $a<|{\bf r}_{s}-{\bf
r}_{t}|<ae^{dl}$ 
can be expanded in $a/|{\bf r}_{i}-{\bf r}_{j}|$ (the distance between distant
charges) and yields 
\[ \fl
-2\pi^{2} dl\int_{|{\bf r}_{i}-{\bf r}_{j}|>\tilde{a}} 
\sum_{({\bf n}_{i},{\bf m}_{i});({\bf n}_{j},{\bf m}_{j})}
y_{{\bf n}_{i},{\bf m}_{i}}y_{{\bf n}_{j},{\bf m}_{j}}
(\alpha_{js}\alpha_{ks}-\beta_{js}\beta_{ks}) 
y^{2}_{{\bf n}_{s},{\bf m}_{s}}
G ({\bf r}_{i}-{\bf r}_{k})
\]
where
$\alpha_{js}=g^{ac}N_{j}^{a}N_{s}^{c}+
p^{2}(g^{-1})^{ac}m_{j}^{a}m_{s}^{c},\beta_{js}=p
(n_{j}^{a}m_{s}^{a}+n_{s}^{a}m_{j}^{a})$ and we have defined the composite
electric component of the charges $N_{j}^{a}=n_{j}^{a}+p~M^{ab}m_{j}^{b}$. 
Using the antisymmetry of $M^{ab}$ and reexponentiating this contribution, 
we can check the renormalisability of
the model to order $y^{2}$ as this contribution can be cast into a
contribution to the matrices $g,M$ and the fugacities. We 
obtain the following corrections to $g$ and $M$ to the order $y^{2}$ : 

\begin{eqnarray}\label{rg-g}
&& \partial_{l}g^{ac}=-2\pi^{2} \sum_{{\bf n}_{s},{\bf m}_{s}}
(g^{ab}g^{cd}N_{s}^{b}N_{s}^{d}-p^{2} m_{s}^{a}m_{s}^{c}) y^{2}_{{\bf
n}_{s},{\bf m}_{s}}\\ 
\label{rg-M}
&& \partial_{l} M^{ab}=-2\pi^{2} p
\sum_{{\bf n}_{s},{\bf m}_{s}}
\left[(g^{-1})^{bc}N_{s}^{a}m_{s}^{c}- (g^{-1})^{ac}N_{s}^{b}m_{s}^{c}
\right] y^{2}_{{\bf n}_{s},{\bf m}_{s}}
\end{eqnarray}
\endnumparts

Notice the antisymmetry of the correction (\ref{rg-M}) to $M^{ab}$. We can now
derive the renormalization flow from these scaling equations. 

\subsection{Specific model and charge asymmetry}
 In the following we will restrict our study to the model (\ref{model-specific}). 
Contrarily to the usual case where the
fugacities of  electromagnetic charges are symmetric both in $n$
and $m$ \cite{nienhuis87}, here this symmetry is broken by the
presence 
of the $\theta$ coupling. As seen above with the T symmetry, changing $n$ to
$-n$ amounts to change also $\theta$ to $-\theta$, and similarly with the
transformation $m\to -m$. The only symmetry which does not modify either $g$ or
$\theta$ corresponds to $(n,m)\to (-n,-m)$, which ensures the neutrality of
the gas for any value of the couplings. Hence we cannot 
assume the $m$ or $n$ parity of the fugacities 
$y_{n,m}$ as the condition $y_{n,m}=y_{-n,m}=y_{n,-m}$ will not 
be 
preserved by the RG.  

For the model (\ref{model-specific}) we need only to consider charges satisfying 
$n^{2}=0,m^{1}=0$. We will use the notation $(n,m)$ for $(n^{1},m^{2})$. 
The above RG equations (9) 
can then be written in a more pleasant way, using the
notation $x=pg^{-1}$ and $t=p~\theta/2\pi$ :
\numparts
\begin{eqnarray}\label{rg2-x}
&& \partial_{l} x =  2\pi^{2} p \sum_{n,m}
\left[ (n+tm)^{2} -x^{2}m^{2} \right] y_{n,m}^{2}+ {\cal  O} (y^{3})\\
\label{rg2-t} 
&& \partial_{l} t
= - 4\pi^{2}p~ x \sum_{n,m}m (n+tm) ~
y_{n,m}^{2}+ {\cal  O} (y^{3})\\\label{rg2-y}
&& \partial_{l} y_{n,m}=2\left(1-\frac{p}{4x} (n+tm)^{2}
-\frac{p}{4}xm^{2} \right) y_{n,m}
+ {\cal  O} (y^{2}) 
\end{eqnarray}
\endnumparts

 In the last equation the non linear term from (\ref{rg-M}) has been
forgotten, being subdominant around each transition point. However it has to
be taken into account when analyzing the topology of the whole phase
diagram. The above RG equations correspond to the starting point of our
analysis of the different transitions. Their invariance under the duality
transformations is made more explicit if we write them as a single equation
for the complex coupling constant $z$ :  
\begin{equation}\label{rg-z}
\partial_{l} z = i 2\pi^{2}p \sum _{n,m}\left(n+zm \right)^{2}y_{n,m}^{2}
\end{equation}
 On this last equation, the invariance under duality of the flow
($z\rightarrow -z^{-1}$) is obvious.  

\begin{figure}[thb]
\centerline{\fig{10cm}{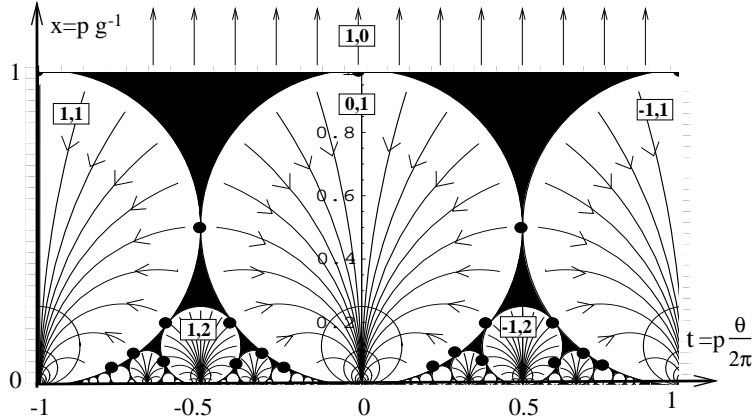}}
\caption{\label{figure1} RG flow in the plane
$p\frac{\theta}{2\pi},pg^{-1}$. The flow are 
shown up to $m=3$. The regions inside the circle corresponds to stability
regions for charges $(n,m)$ with nonzero $n$, 
$m$ labelling the size of the circle while $n$
gives their positions. For $pg^{-1}>1$ only charges with $m=0$ exists. The
transition points between $m\neq 0$ phases are shown as black points while the
black region corresponds to a phase with no charge (neutral phase).}
\end{figure}

 The invariance under duality of these beta functions, which was assumed in
previous studies of modular invariant models, have important
consequences : as the axis $t=0$ is invariant (in the limit of vanishing
fugacities) and thus correspond to a flow line of the above equations, we know
from the action of the modular group on this flow line that all the flow lines
of the phase diagram are either straigth lines or arcs of circles. 
The flow diagram obtained by numerically
integrating these equations is shown on figure \ref{figure1}, where only
charges with $m\leq 3$ have been taken into account. 

\subsection{Analysis of the flow}\label{part:analyse}
 
 To derive the phase diagram from the renormalization analysis, we first find
the domains where a charge $(n,m)$ proliferate. Such a charge proliferate,
or equivalently the parafermion operator (\ref{parafermions}) associated with
a charge $(n+\frac{\theta}{2\pi}m,m)$ is
relevant, when the associated fugacity increases under rescaling. From the
renormalization eigenvalue of the fugacities in (\ref{rg2-y}) we deduce that  
the pure electric
charges $(n,0)$ are relevant only for $x>pn^{2}/4$. For $x<1$ 
the composites charges $(n,m)$ become relevant 
inside the circles $[n,m]$ centered on $(t,x)= (-n/m,2/(pm^{2}))$ of radius 
$2/(pm^{2})$. All this circles are thus tangent to the axis $x=0$ (see
figure \ref{figure1}). 

 Each circle [n,m] is caracterised by a ratio $\nu=n/m$ instead of $n,m$, as
all charges $(\lambda n,\lambda m)$ with $\lambda\in \mathbb{Z}$ are generated under
renormalization 
in this circle (see figure \ref{figure1}). 
In the following we will write
$\nu=n/m$ where $m$ is the minimal magnetic charge allowed in the circle. 
 Under renormalization, in a circle $\nu=n/m$, $x=pg^{-1}$ flows to zero while
$t$ is renormalized to $t^{*}=\nu$. Hence each rational point of the axis
$x=0$ correspond to a fixed point of the RG, caracterising a given phase. 
 The phase $m=0$, stable for $x>pn^{2}/4$, can be viewed as a special circle of
infinite radius, caracterised by $\nu=\infty$. In this phase $x\to \infty $
and $t$ is slightly renormalized to a (non universal) real value. Below
$x<pn^{2}/4$ and between all these circles, all the fugacities renormalize to
zero, which correspond to a neutral phase of the Coulomb gas, caracterised by
non universal renormalized $g$ and $\theta$ and vanishing fugacities.  

 Transitions in this phase diagram are a priori of two different types : the
transitions between two circles caracterised by different ratio $\nu$
and the transition between a phase with relevant charge $(n,m)$ and the
neutral phase. Transitions of the first kind correspond to tangent points
between circles  in the phase diagram. Such points exist only when $p=4$. For
$p<4$, the circles overlap and these transitions are no more accessible by the
present perturbative study, while for $p>4$ all circles with different ratio
$\nu$ are disconnected and these transitions disappear.  In the following we
will consider in more detail the case $p=4$, whose phase diagram is shown on
figure \ref{figure1}.

 A transition between two phases (ratios) 
$\nu_{1}$ and $\nu_{2}$ is allowed if the two
corresponding circles are tangent to each other, which can be written    
\begin{equation}\label{relation-transition}
\nu_{1}-\nu_{2}=\pm \frac{1}{m_{1}m_{2}}
\end{equation}
 The corresponding transition point is located in 
$(t,x)= (- (n_{1}m_{1}+n_{2}m_{2})/
(m^{2}_{1}+m_{2}^{2}),1/(m^{2}_{1}+m_{2}^{2}))$. To analyze further these
transitions, we need to remark that any circle $\nu$ can be deduced from the
circle $\nu=0$ by succesive application of the transformations $P,P^{-1}$ and
the duality $D$. Moreover the circle $\nu=0$ itself is the image under $D$ of
the line of phase transitions $x=1$. The transition points between the phase
$\nu=0$ and other $\nu'=\pm 1/m$ (see figure \ref{figure1}) are all images
under $D$ of the
transitions points between $[1,0]$ and the phases $[n,1]$, which themselves
are related to the transitions $[1,0]\leftrightarrow [0,1]$ by action of $P$
and $P^{-1}$. Thus all the transition points between two ratios $\nu$ and
$\nu'$ correspond to the same critical behaviour. Similarly all transitions
between a phase $\nu$ and the neutral phase (no charge condensate) are images
of the transition between $[1,0]$ and this phase. We can now study these two
transitions perturbatively in the $y_{n,m}$.    

 First at a transition between $\nu_{1}=n_{1}/m_{1}$ and $\nu_{2}=n_{2}/m_{2}$
 we can write renormalization equations for the distance orthonormal to the
transition point $\delta_{12}$ while the distance perpendicular to this axis
 is not renormalized to lowest order. Using $\epsilon=
(m_{1}^{2}+m_{2}^{2})\delta_{12}$ and
$y=y_{n_{1}m_{1}},\tilde{y}=y_{n_{2}m_{2}}$ we obtain 

\[
\partial_{l} \epsilon=2\pi^{2} (y^{2}-\tilde{y}^{2})
\quad \quad \quad \quad \partial_{l}y=2\epsilon y
\quad \quad \quad \quad \partial_{l}\tilde{y}=2\epsilon \tilde{y}
\]
 which correspond to a Kosterlitz-Thouless like diverging correlation length
$\xi\sim \exp (cte/|\epsilon|^{\frac{1}{2}})$. In both phases, the
correlation functions of the parafermion operators decays exponentially : 
\[
\left< 
e^{i (n\phi ({\bf r})+m\tilde{\phi} ({\bf r}))} 
e^{-i (n\phi ({\bf o})+m\tilde{\phi} ({\bf o}))} 
\right>\sim 
e^{-\frac{r}{\xi}}
\]

 The transition between the neutral phase and a circle $\nu$ is describe by a
critical point with one marginal direction (tangent to the circle) and the
same diverging correlation length. While the correlation function have the
same kind of exponential decays from the circle side of the transition, they
decay algebraically in the neutral phase :    
\[
\left< 
e^{i (n\phi ({\bf r})+m\tilde{\phi} ({\bf r}))} 
e^{-i (n\phi ({\bf o})+m\tilde{\phi} ({\bf o}))} 
\right>\sim 
\left(\frac{a}{r} \right)^{\frac{1}{x} (n+tm)^{2}+xm^{2}}
\]

\section{Analogy with the Quantum hall effect}

Within the context of the two parameters scaling theory of the 
quantum hall effect,
the first idea consists in identifying the coupling of the Coulomb gas model with
the components of the conductivity tensor of the electron gas  
$\sigma_{xy}=t,\sigma_{xx}=x$. The ratio $\nu$ defined in the above study
naturally translate into the filling factors of the quantum Hall states : each
circle of the flow diagram correspond to a given quantum Hall fluid  and
 the renormalization equations we obtain 
provide the expected asymptotic value for this
conductivity : $\sigma_{xy}^{*}=\nu, \sigma_{xx}^{*}=0$. Following this
analogy we can find that
transitions between two filling factors $\nu_{1},\nu_{2}$ are allowed if they
satisfy the expected relation (\ref{relation-transition}).   All the allowed
transitions between two plateaux are in the same universality, being related
by the duality D (\ref{dual}) which exchange the electric and magnetic
charges, as in \cite{shahar96}, thus supporting the superuniversality
hypothesis\cite{kivelson92}.  

However we notice that in our flow diagram, filling factors with any
denominators exist, while only even denominators are expected for quantum hall
fluids. Moreover fixed points corresponding to filling factors (circles) with
even and odd denominators  are related by modular transformations : hence one
cannot avoid the even denominators in this model. This result has to be
related with recent works \cite{georgelin97} 
which reveal that the symmetry hidden behind the quantum
Hall hierarchies 
should be the subgroup $\Gamma (2)$ of
$SL (2,\mathbb{Z})$ instead of the whole modular group. This subgroup is
generated by the transformation $z\to -z/ (2z+1)$ (instead
of $z\to -z^{-1}$ in this study) which preserves the oddness of filling
factors. This transformation, when restricted to the axis $t=0$ (imaginary axis),
 correspond to $g\to g/ (1+4g^{2})$. Hence it does not exchange the high and
low temperature phases of the usual 
electromagnetic Coulomb gas\cite{kadanoff78}. Thus an extension of the usual
electromagnetic model to a $\Gamma (2)$
invariant theory will not simply correspond to the addition of a topological
term as for the model studied in this article. 
 
 To conclude, let us comment on the comparison with the recent studies which
focus on the constraint on the beta function from the modular symmetry
\cite{dolan98}. This study (together with \cite{burgess97})
 assume a modular (or $\Gamma (2)$) symmetry in a model with two parameters
$\sigma_{xx},\sigma_{xy}$, and derive the general form of the beta functions
of this model. In our work the model (\ref{partition}) possess an infinite
number of coupling constants corresponding to the electromagnetic charge
fugacities. Thus a direct comparison between the beta functions is not
possible. However it should be noticed that our renormalization procedure
provide the first explicit example of the commutation of the modular symmetry
and the renormalization, which is the central hypothesis of
\cite{burgess97,dolan98}.

\section{Conclusion}

 In this article we thus extended the renormalization study of
\cite{nienhuis87} to the Coulomb gas with a topological $\theta$ term in two
dimensions. The renormalization scheme we used provides scaling equations
which were found to be themselves invariant under the modular group. 
To our knowledge
this study provide the first example in two dimensions of an explicit
renormalization of a modular invariant model.   
 The rich phase diagram of this model certainly deserves more work : of particular
interest will be the applications of this renormalization study to the
extensions of the various statistical models, related to the electromagnetic
Coulomb gas in the 
absence of the $\theta$ term.

\bigskip 
{\it Note added} : after completion of this work, a preprint
\cite{cristofano98} appeared on a
related quantum Hall model where a finite electromagnetic background charge is
added to the 
$k=1$ restriction of the model (\ref{action}) . Although these authors used a
simplified scheme, their scaling analysis seem to agree with the complete RG
equations we derived in this article. 

\ack
 I thank Benoit Dou{\c c}ot for pointing out to me reference \cite{callan91}. 
\vskip-12pt
\section*{References}



\end{document}